\documentclass[aps,prd]{revtex4}

\newcommand{\be}{\begin{equation}}
\newcommand{\ee}{\end{equation}}
\newcommand{\rf}[1]{(\ref{eq:#1})}

\usepackage{graphicx}

\begin{document}

\title{Perturbations of  Dark Matter Gravity}

\author{M. D. Maia\thanks{maia@unb.br},
A. J. S. Capistrano\thanks{capistranoaj@unb.br} \& D.
Muller\thanks{muller@fis.unb.br}\\
{Universidade de Bras\'{\i}lia, Instituto de F\'{\i}sica,
Bras\'{\i}lia, DF.70919-970}}

\begin{abstract}
Until  recently  the  study   of the gravitational  field of  dark
matter was  primarily   concerned   with its  local  effects on the
motion of   stars  on galaxies  and   galaxy clusters.  On the other
hand,  the  WMAP   experiment   has   shown that    the
gravitational  field produced by dark matter  amplify  the higher
acoustic modes  of the  CMBR   power  spectrum, more  intensely
than  the   gravitational  field  of  baryons.   Such  wide  range
of  experimental  evidences from cosmology  to  local  gravity
suggests   the necessity of  a  comprehensive   analysis of the
dark matter gravitational  field per se, regardless  of any other
attributes that  dark matter may  eventually possess.

In the present  note  we introduce and   apply   Nash's  theory  of
perturbative  geometry  to the  study  of  the dark matter
gravitational field  alone, in a  higher-dimensional framework. It
is  shown that    the dark matter gravitational perturbations in the
early  universe   can be  explained by   the  extrinsic  curvature
of  the  standard  cosmology. Together  with  the estimated
presence of  massive  neutrinos,   such geometric perturbation is
compatible not only  with  the observed  power  spectrum in  the
WMAP  experiment, but  also   with the most recent  data  on  the
accelerated  expansion of the universe.

It is possible that  the  same  structure  formation   exists  locally,   such  as in  the cases of  young galaxies  or in  cluster  collisions. In most other cases  it  seems    to have ceased, when  the extrinsic  curvature becomes  negligible,  leading  to  Einstein's  equations
in  four-dimensions.   The slow motion of  stars   in  galaxies and
the    motion of plasma  substructures  in nearly  colliding clusters,  are calculated  with the geodesic  equation for  a  slowly moving   object  in a gravitational  field  of  arbitrary strength.
\end{abstract}

\maketitle

\section{Dark Matter Gravity}
The     dark matter      concept   originated  from the
 observed  discrepancy between the measured  rotation velocity curves
for stars in a spiral galaxy and  clusters  and the theoretical
prediction from the  Newtonian gravitational theory. This was  first
noted   by F. Zwicky, when looking at the Coma cluster in 1933
\cite{Zwicky}. The  measurement  of the   velocities  are  based on
the  Tully-Fisher relation between the mass of the galaxy and the
width of the 21-cm line of hydrogen emissions \cite{Tully}. As
figure 1 shows, the observed curve becomes almost horizontal (flat),
unlike that produced by the Newtonian theory \cite{Albada}. Similar
patterns  occurs   in most spiral galaxies and galaxy clusters
\cite{Yoshiaki}.
\begin{figure}[!h]
\begin{center}
\includegraphics[width=6cm]{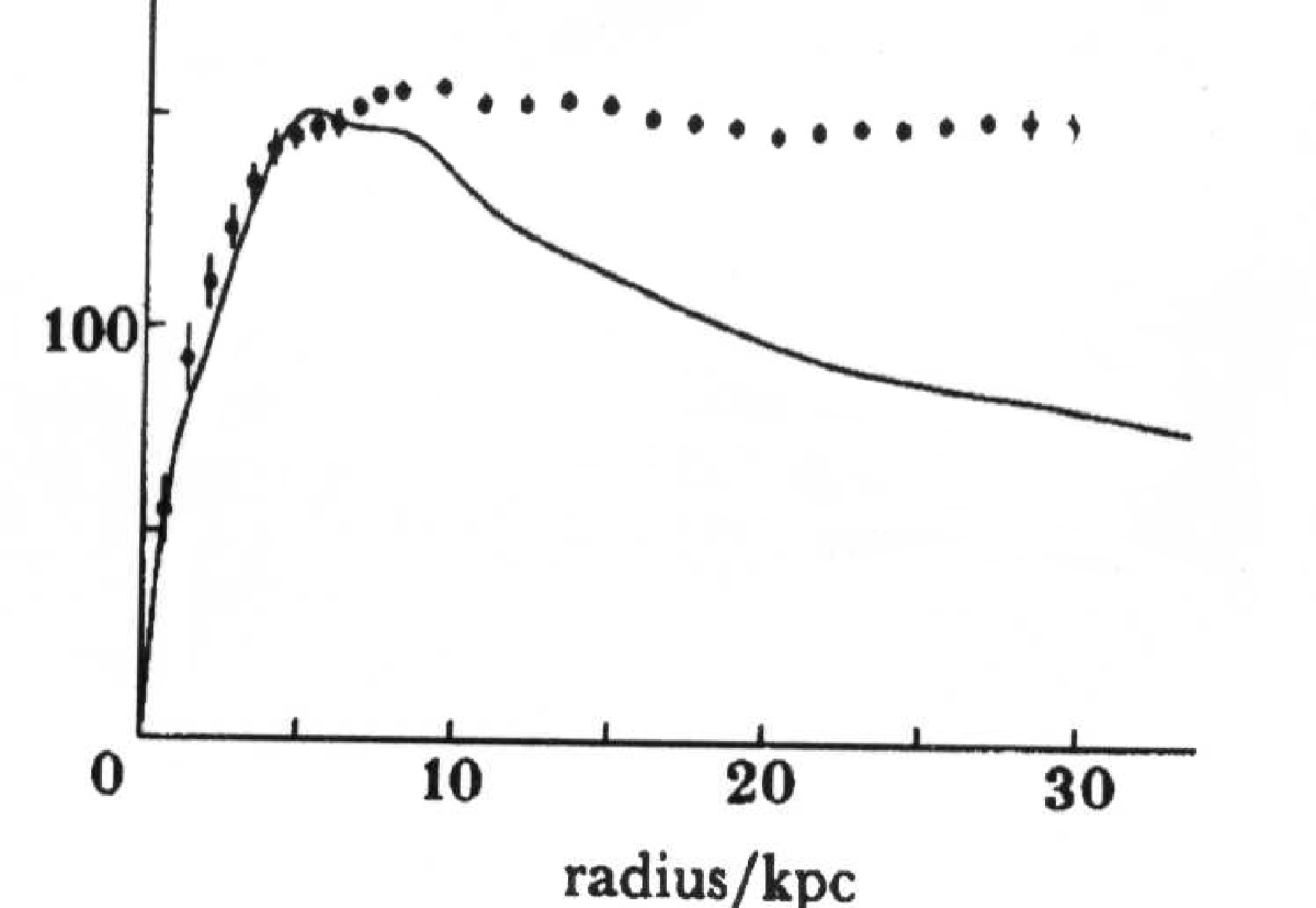}
\caption{ \footnotesize Observed Rotation velocity curve of the
NGC3198 (error bars) and the prediction from Newtonian theory.}
\end{center}  \label{fig:albada}
\end{figure}
Except for the  fact  that  dark matter  interacts mainly through its  gravitational  field,  not much is  know about  its
other physical properties. This  has  motivated    several  attempts
to    dismiss   dark matter   altogether,  regarding  its
gravitational  effects  as   evidences  for  an  alternative
gravitational  theory
with respect to Newtonian theory\cite{Milgrom,Moffat}, or for  variants   of general relativity\cite{Bekenstein,Freese,Skordis}.
General relativity itself has been  traditionally excluded  from  this   analysis essentially
because the     slow  motion of  the observed  objects  usually
leads  to   the  Newtonian  limit  of the theory. This is
reinforced by the fact  that   near the galaxies  nuclei  where the
gravitational field is strong, like that of  a black hole, the observed  velocities closely
agree  with   the prediction from Newton's theory (as in Figure 1.),
leading  to   the conclusion that  Newton's gravity should apply
everywhere else,   but  a   sufficient amount of    dark matter must
be  added  to  increase  the Newtonian gravitational pull.
A recent   review  and  check list   for  dark matter candidates  can  be  found in \cite{Hooper,Marco}.

A  qualitative   distinction  between   dark matter gravity   and
baryons gravity was evidenced in  the  WMAP  experiment \cite{Spergel}.  Referring to   Figure 2,  we  quote  :
 \emph{..."Cold dark matter serves as a significant forcing term that amplifies the higher acoustic oscillations. Alternative gravity models (e.g., MOND), and all baryons-only models, lack this  forcing term so they predict a much lower third peak than is  observed by WMAP and small scale CMB experiments"...}\\
\begin{figure}[!h]
\begin{center}
\includegraphics[width=10cm]{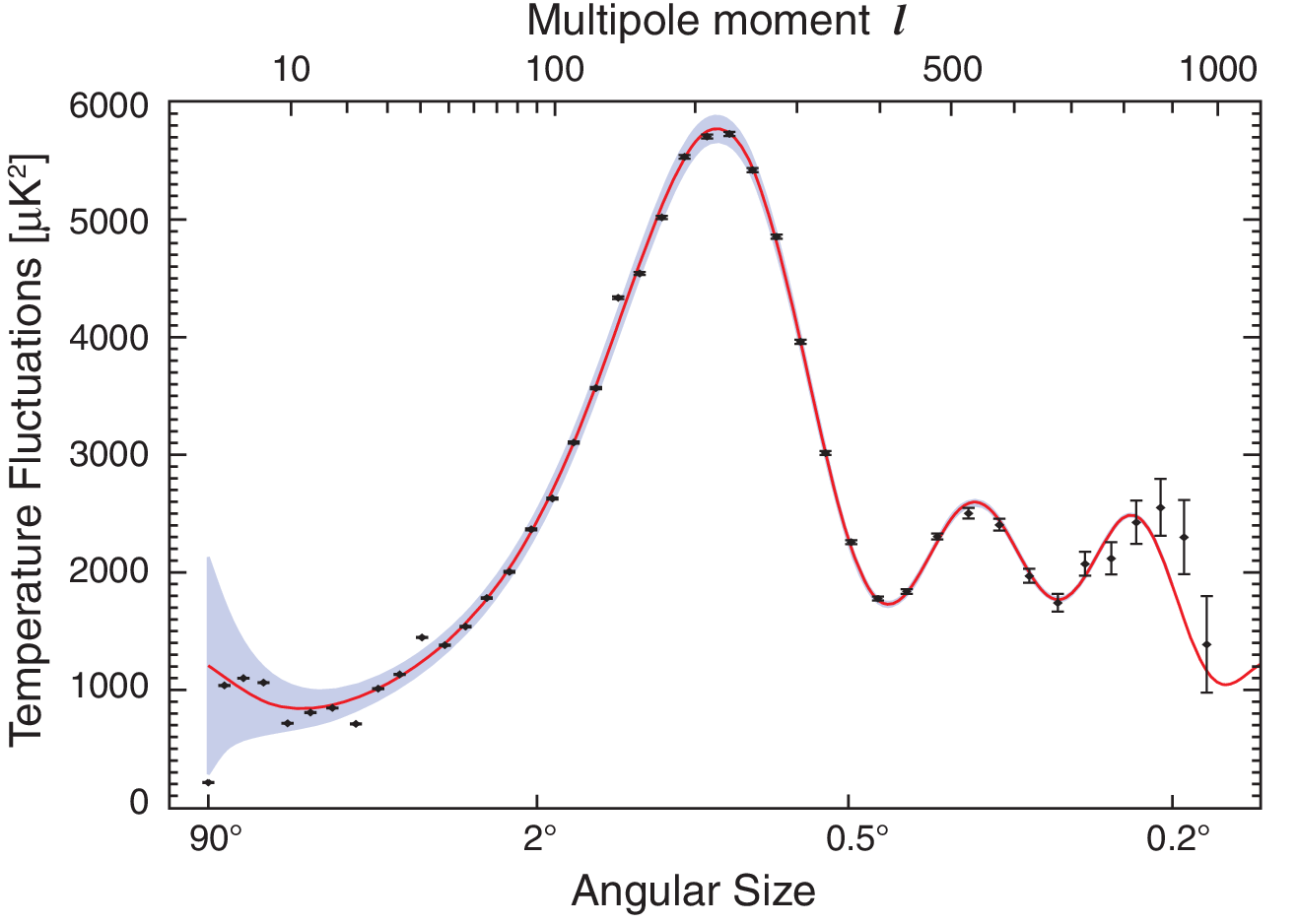}
\caption{ \footnotesize The observed   power  spectrum of  the CBMR
of the  WMAP experience shows  how   the  dark matter gravitational
field  can be   different from  the baryonic  one.  }
\end{center}\label{Fig:wmap}
\end{figure}
 These results were confirmed and sharpened by  the observations of the luminous red galaxies in the Sloan Digital Sky Survey (SDSS), supporting the $\Lambda CDM$ model \cite{Tegmark}.  This  was   improved in the  recent   WMAP fifth
 year  report  \cite{WMAP5}.

As it  seems   evident,  the  study of dark matter gravity  and its
implications to the formation of  structure  in the early universe
must  naturally  start   with gravitational perturbation theory \cite{Rocky,Mukhanov,Padmanabham}.
The  traditional gravitational   perturbation mechanisms in
relativistic  cosmology  are  plagued  by coordinate  gauges, mostly
inherited  from the group of diffeomorphisms  of  general
relativity. Fortunately   there  are some     very  successful
criteria to  filter out the latter perturbations
\cite{Bardeen,Geroch,Walker},   but they still depend on  a  choice
of a perturbative  model.
 A less known, but  far   more general  approach to
gravitational perturbation  can be derived  from  a  theorem  due to
  John   Nash,   showing   that   any  Riemannian   geometry    can be  generated by  a    continuous  sequence  of local  infinitesimal   increments of a  given  geometry \cite{Nash,Greene}.

Nash's  theorem   solves    an old  dilemma     of  Riemannian
geometry,  namely that the Riemann tensor   is not sufficient  to
make  a precise  statement    about  the local shape  of  a
geometrical object or    a manifold. The simplest   example is
given  by a 2-dimensional Riemannian  manifold,  where  the Riemann
tensor  has only  one  component $R_{1212}$  which  coincides with
the  Gaussian  curvature. Thus,   a  flat Riemannian    2-manifold
defined by  $R_{1212}=0$   may be interpreted as a plane,  a
cylinder or a   even  a helicoid,  in the sense of  Euclidean
geometry. Riemann regarded   his  concept of  curvature  as defining an  equivalent class of  manifolds instead of  an  specific  one \cite{Riemann}.
While  such  equivalence of  forms is  mathematically interesting,  it  is  less than  adequate     to    derive physical  conclusions
from today's  sophisticated   astronomical observations.

The solution to  Riemann's  ambiguity problem  was originally
proposed in  1873 by  L.  Schlaefli \cite{Schlaefli}, conjecturing
that if a  Riemannian  manifold   could be  embedded in  another
one, then   a decision  on its  real  shape  could  be  made  by
comparing  the Riemann  curvatures of the embedded   surface  with
the one  of the embedding  space. The  formal  solution of this
problem  took a long time  to be  developed  and it  came only
after the  derivations of the  conditions  that guarantee  the
embedding of any Riemannian  geometry into another,   the   well
known Gauss-Codazzi-Ricci  equations of  geometry.    The  most
general  solution  of the Schlaefli's  conjecture appeared  only in
1956 with  Nash's theorem.

The   Gauss-Codazzi-Ricci   equations  are  non-linear and
difficult to  solve  in the general case. Some   simplifications
were   obtained by  assuming that the   metric is    analytic in the
sense that it  is  a convergence of  a  positive power  series
\cite{Cartan,Janet}. Nash's  theorem innovated the   embedding
problem  by  introducing the  notion of  differentiable,
perturbative geometry:  Using  a  continuous sequence  of  small
perturbations  of a  simpler  embedded  geometry along  the  extra
dimensions, he  showed  how to construct  any other  Riemannian
manifold\footnote{To the best of  our knowledge, the geometric
perturbation method   was first  introduced  by  J. Campbell in
1926, in  a posthumous edition of his textbook  on differential
geometry \cite{Campbell}. Unfortunately, the relevance of  the
perturbative process  was spoiled   by  the use of analytic
conditions\cite{Dahia}.}\cite{GDE,QBW}.  Nash's
approach  to    geometry not  only  solves the ambiguity  problem
of the  Riemannian   curvature, but also  gives  a prescription on
how to  construct geometrical  structures by gradual   deformations
of   simpler ones.

The purpose of this  paper  is  to  show  that    Nash's  geometric perturbative process  contributes  to   explain the formation  of  structures  in the universe.  To see  how this works, in the  next  section  we  show
how  the geometry of the  standard   cosmology regarded as
submanifold embedded a  5-dimensional  deSitter bulk, can be
perturbed  a la Nash,  leading  to  a  modified  Friedman's
equation.
 We  draw the   theoretical  CMBR  power  spectrum resulting from  such  perturbation,   comparing it  to   the  observed spectrum in the  WMAP  experiment. In section 3   we  apply the same  procedure   for  local  dark matter gravity. However, since there are  no evidences  that  the   perturbative  process is  still going on  today,  except  perhaps  in  young  galaxies and  cluster  collisions  where  experimental data  is  still  scarce,   we  assume  as  a  first     estimate  that   Nash's perturbation  is  locally   negligible  in most  spiral galaxies. This leads   to  a  simpler  set of  equations  where  the
brane-world  gravitational  equations reduce  to the  usual
four-dimensional  Einstein's  equations. In this case  we  will  see
that the     motion  of  stars  and  of  plasma  substructures  can
be    described  by the   application  of  "slow  geodesic motion"
equation  which  holds true  in  an gravitational field  of
arbitrary  strength and is  not  limited  to the weak gravitational
fields  as in  General Relativity.

\section{Dark Matter Cosmology}

We start  by reviewing some   basic  ideas of   Nash's  geometric
perturbation
 theorem:  Suppose  we have  an  arbitrarily   given   Riemannian  manifold  $\bar{V}_n$ with metric   $\bar{g}_{\mu\nu}$,  which  is  embedded  into  a given
higher dimensional  Riemannian  manifold  $V_D$,  the bulk  space.
Then  we may  generate  another  metric  geometry  by  a     small
perturbation $\bar{g}_{\mu\nu} +  \delta g_{\mu\nu} $
where\footnote{ Greek  indices  $\mu,\; \nu...$ refer  to  the $n$
dimensional  embedded  geometries;  Small case Latin  indices $a, \;
b...$ refer  to $N$ extra  dimensions;  Capital  Latin indices
$A,\; B...$ refer  to  the  bulk} \be \delta g_{\mu\nu}
=-2\bar{k}_{\mu\nu  a}\delta y^a,  \;\; a =N\!+\!1...D
\label{eq:York} \ee where  $\delta  y^a$ denotes  an  infinitesimal
variation of the  extra dimensions orthogonal to  $\bar{V}_n$  and
  $\bar{k}_{\mu\nu a}$  denote the  extrinsic curvature
components of  $\bar{V}_n$ relative to  the extra dimension $y^a$.
 Using this  perturbation we obtain
 new  extrinsic  curvature $k_{\mu\nu  a}$, and  by repeating  the process
  we  obtain  a  continuous  sequence of  perturbations like
\[
g_{\mu\nu}  =  \bar{g}_{\mu\nu}  +  \delta y^a \, \bar{k}_{\mu\nu a}  +
\delta y^a  \delta y^b \, \bar{g}^{\rho\sigma}
\bar{k}_{\mu\rho a}\bar{k}_{\nu\sigma b}\cdots
\]
In this  way    any    Riemannian  geometry   can be generated.

Nash's original  theorem   used  a  flat D-dimensional  Euclidean
space  but this  was   soon generalized  to  any  Riemannian
manifold, including those  with   non-positive  signatures
\cite{Greene}.  Although the  theorem  could  also be generalized to
include  perturbations  on  arbitrary  directions in the bulk,    it
would  make  its interpretations  more difficult,   so that we
retain  Nash's   choice of  independent orthogonal  perturbations.
It  should be noted that  the   smoothness of the embedding is
a    primary  concern   of Nash's   theorem. In this  respect, the
natural  choice  for  the  bulk is that  its  metric  satisfy the
Einstein-Hilbert  principle. Indeed,  that principle represents  a
statement  on   the  smoothness  of the embedding  space (the
variation  of the Ricci scalar  is the  minimum possible).
Admitting that the perturbations are  smooth (differentiable),  then
the  embedded  geometry  will  be  also differentiable.

 The Einstein-Hilbert principle leads  to the  D-dimensional Einstein's  equations  for the  bulk
metric  ${\cal G}_{AB}$  in  arbitrary  coordinates
 \be
{\mathcal{R}}_{AB}-\frac{1}{2}\mathcal{R} \mathcal{G}_{AB}
 =\alpha_* T^*_{AB}\label{eq:EEbulk}
\ee
 where   we  have dispensed  with    bulk  cosmological  constant and  where $T^*_{AB}$ denotes the energy-momentum tensor   of the known  matter and gauge fields. The  constant  $\alpha_*$ determines the D-dimensional energy scale.

The  four-dimensionality of the    space-time manifold  is  an
experimentally  established fact,  associated  with  the   Poincaré
invariance of  Maxwell's  equations and  their  dualities, later
extended to all gauge fields. Therefore,   all matter  which
interacts   with these gauge  fields  must  for  consistency  be
also   defined in  the  four-dimensional  space-times.  On the other
hand,  in  spite of  all  efforts  made so far, the gravitational
interaction  has failed to  fit into  a  similar  gauge  scheme,  so
that the  gravitational  field  does  not necessarily have the same
four-dimensional  limitations and it   can  access  the extra
dimensions in accordance  with \rf{York},  regardless  the  location
of  its  sources.

 We  assume  that the  four-dimensionality   of   gauge   fields  and  ordinary matter  applies  to all perturbed  space-times,  so  that  it  corresponds  to   a  confinement  condition.  In order  to   recover  Einstein's gravity  by reversing the  embedding, the   confinement  of  ordinary matter and  gauge  fields implies that
the  tangent  components of  $\alpha_* T^*_{AB}$  in the  above
equations must  coincide  with  $8\pi G  T_{\mu\nu}$ where
$T_{\mu\nu}$  is the   energy-momentum  tensor of the confined
sources \footnote{ As it may   have  been already noted,   we  are
essentially  reproducing the  brane-world  program,  with the
difference that  it is  very  general  and  it  has  nothing to  do
with branes in string/M theory. Instead,  all  that we  use here is
Nash's   theorem  together  with  the  four-dimensionality of gauge
fields,  the  Einstein-Hilbert principle for the bulk and  a
D-dimensional energy scale $\alpha_*$. }.

 Since  dark matter gravity  is not  necessarily confined, it  also propagates  in the bulk. Furthermore, based on the lack of  experimental evidences on the energy-momentum of the dark matter in the bulk,  we also assume  that the normal  and  the  cross components of the  dark matter  energy-momentum  tensor, respectively   $T_{\mu a}$ and  $T_{ab}$  vanish, meaning  that there  are  no  known  sources  outside  the  four-dimensional  space-times.

The   standard   Friedman-Lemaitre-Robertson-Walker (FLRW) universe
can be  embedded without  restrictions   in  a five-dimensional.  Therefore, using  the  Einstein equation \rf{EEbulk}  written
in the Gaussian  frame defined by the  four-dimensional submanifold, we obtain the  equations of the FLRW   embedded  geometry\cite{GDE}
 \begin{eqnarray}
&&R_{\mu\nu}-\frac{1}{2}Rg_{\mu\nu} -
{Q}_{\mu\nu} = -8\pi G T_{\mu\nu} \label{eq:BE1}\\
 &&k_{\mu ;\rho}^{\rho}\!
 -\!h_{,\mu} = 0\label{eq:BE2}
\end{eqnarray}
where now $T_{\mu\nu}$ is the  energy-momentum tensor of the
confined perfect  fluid,  $k_{\mu\nu 5} \equiv k_{\mu\nu}$ denotes
the  components of the extrinsic curvature of the   embedded
space-time,  $h= g^{\mu\nu}k_{\nu\nu}$, $K^{2}=k^{\mu\nu}k_{\mu\nu}$
and
 \begin{equation}
Q_{\mu\nu} = k^{\rho}{}_{\mu }k_{\rho\nu }-h
k_{\mu\nu}\!\!-\!\!\frac{1}{2}(K^{2}-h^{2})g_{\mu\nu}\label{eq:Qij}.
\end{equation}
This  tensor  is  independently conserved, as it can be  directly
verified that (semicolon  denoting  covariant  derivative  with
respect  to  $g_{\mu\nu}$) \be Q^{\mu\nu}{}_{;\nu} =0
\label{eq:cons} \ee In coordinates $( r, \theta, \phi , t )$ the
FLRW model can be expressed s \cite{Rosen}: \be
ds^{2}=-dt^{2}+a^{2}(t)[dr^{2}+f_k^{2}(r)(d\theta ^{2}+
\mbox{sin}^{2}\theta d\phi ^{2})]\label{eq:FLRW} \ee
 where $f_k(r) = r, \mbox{sin}\, r, \mbox{sinh}\, r $
corresponding to $k =0, +1, -1$ (spatially flat, closed, open
respectively). We   start  solving    \rf{BE2}   for the  above
metric in the deSitter bulk. It is  easier to  find  the  solution
of   Codazzi's equations
\[
k_{\mu[\nu;\rho]}  =0
\]
of   which   \rf{BE2}  is  just  its  trace. The  general  solution
of  this  equation is
 \be
 k_{ij}=\frac{b}{a^{2}}g_{ij},\; \;\,
 k_{44}=-\frac{1}{\dot{a}}
\frac{d}{dt}\left(\frac{b}{a}\right)\;\; i,j = 1\ldots
3\label{eq:kij}
 \ee
 Defining $B=\frac{\dot b}{b} $, we may express the  components of  $Q_{\mu\nu}$  as
 \begin{eqnarray}
\label{eq:BB}
&&Q_{ij}= \frac{b^{2}}{a^{4}}\left( 2\frac{B}{H}-1\right)
g_{ij},\;\;\;Q_{44} = -\frac{3b^{2}}{a^{4}},\; \;\; Q=
-\frac{6b^{2}}{a^{4}} \frac{B}{H},  \;\;\;i,j =1..3
 \end{eqnarray}
where $H= \dot{a}/a$ is the usual Hubble  parameter. After replacing
in \rf{BE1}  we obtain Friedman's equation   modified by  the
extrinsic  curvature:
\begin{equation}
(\frac{\dot{a}}{a})^2+\frac{k}{a^{2}}=\frac{8\pi G}{3}\rho
+ \frac{b^2}{a^4} \label{eq:Friedman}
\end{equation}

To  interpret  this  result  we  have compared it  with  the XCDM,
phenomenological   x-fluid model,  with  state equation $ p_{x} =
\omega_{x} \rho_{x}$,   which  corresponds  to the  geometric
equation  on $b(t)$ \be
 \frac{\dot{b}}{b}=\frac{1}{2}(1-3\omega_{x})\frac{\dot{a}}{a}\label{eq:eforb}
 \ee
This  cannot be readily  integrated   because   $\omega_x$ is not
known. However, in  the   particular  case when
$\omega_{x}=\omega_{0}$=constant, we obtain  a simple solution
 \be
 b(t) =
b_{0}(\frac{a}{a_{0}})^{\frac{1}{2}(1-3\omega_{0})} \label{eq:b}
 \ee
 where $a_{0}$ and $b_{0} \neq 0$ are integration constants.
  Replacing  this   solution in  \rf{Friedman}  we obtain  an  accelerated  universe  which is  consistent  with  the most recent observations,  when  the  values of $\omega_0$    are taken  within the range   $-1\leq \omega_{0} \leq -1/3$ \cite{GDE}. Furthermore, the theoretical  power  spectrum   obtained  from the  extrinsic  curvature perturbation of the  FLRW model,    is    not  very
different  from  the  observed  power  spectrum  in the  WMAP/5y,
displayed in Fig.  2.
\begin{figure}[!h]
\begin{center}
\includegraphics[width=6cm]{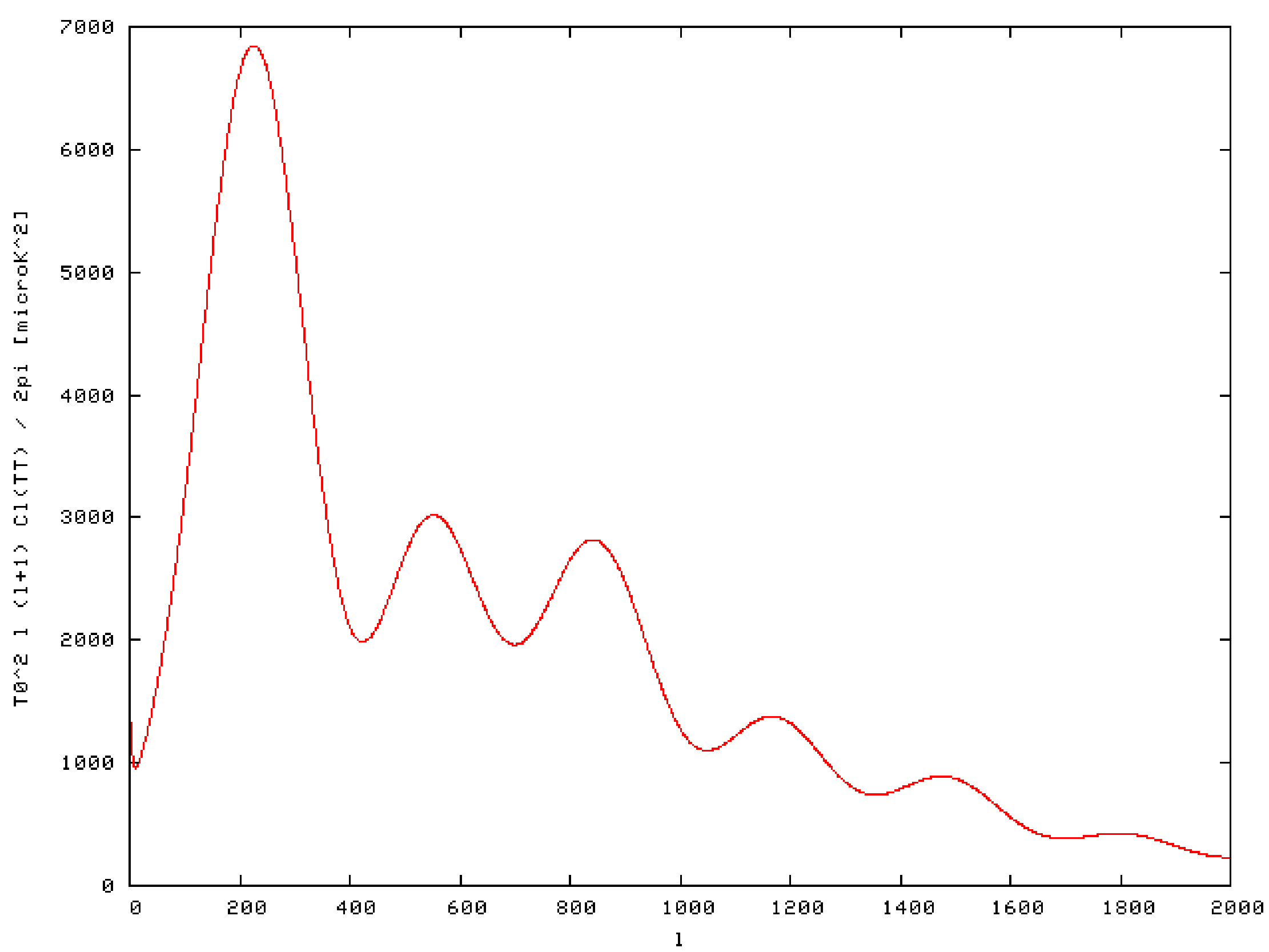}
\caption{\footnotesize The theoretical  power  spectrum calculated
with the  CAMB  for  $   -1 \le  \omega_0 \le -1/3$, Massive
Neutrinos=1, massless  neutrinos  =3.04.}
\end{center}  \label{fig:CAMB01}
\end{figure}

\newpage

\section{Local Dark Matter Gravity}

It is possible that in   young galaxies which are  still in the
process of  formation \cite{Puzia};   in  galaxies    with   active
galaxy  nuclei  \cite{deVries};   or  even  in  cluster  collisions,
Nash's perturbations    could  be
applied,  where the metric  symmetry  is  taken  to be local.
However, there  not    sufficient  experimental  data    to  support
such local perturbations.  Therefore,  in the  following  we
restrict  our study on  local dark matter  gravity  to the cases
where  Nash's perturbations   has  ceased.  From \rf{York}  we  see
that this  limiting  condition is   given by  $k_{\mu\nu}$.  Furthermore, if $T_{\mu\nu}$
is   composed by  ordinary  matter  and  gauge  fields, the rsulting equations  are  the same   as the vacuum  Einstein's  equations
\be
R_{\mu\nu}-\frac{1}{2}Rg_{\mu\nu}=0  \label{eq:EEvac}
\ee
These  equations  are  be understood  in the context of  the  embedded
space-times and  with the  confinement  conditions for  ordinary
matter  and gauge fields. They  do not represent  the  whole of
General Relativity  because the principle of   general  covariance
does not necessarily apply  to the bulk geometry. This  follows
from  the fact  that   Nash's  perturbations  are  restricted to be
along  the orthogonal  directions   only.


The   recently observed  correlation  between     the highest-energy cosmic  rays  sources and  active galactic   nuclei\cite{Auger}
suggests  the existence of  strong gravitational  fields in those  regions\cite{Richstone,Ferrarese1,Ferrarese2}. Nonetheless,   the
observed   motion  of   stars  near black holes observed at the  nuclei  of  some  of these galaxies indicate that the  velocities  are  small, of the order  of a few  hundreds  of  kilometers per  second.  This  requires the  description of   a  slow  geodesic  motion in the presence of  strong gravitational fields
In  what  follows  we  use  essentially the  first  part  of the exposition in  the  Misner, Thorn \&  Wheeler's  book \cite{Wheeler}.

Consider  a   slow  free  falling particle (or  star) in   a
gravitational field  described by  a  solution of the vacuum Einstein's
equations (like in  \rf{EEvac} in  a suitable   coordinate  system.
Initially the particle   is  located  in a  far away  region,  where
the action of the gravitational field is  supposedly weak,  and   we
can  write (The  notation  $h_{\mu\nu}$   reminds that this is not
Nash's perturbation)
 \begin{equation}
 g_{\mu\nu}=\eta_{\mu\nu} +\delta h_{\mu\nu},\;\;\mbox{where}\;\;
  \delta h^2_{\mu\nu}<<\delta
 h_{\mu\nu},\label{eq:weak}
 \end{equation}
Here the    small deviation   from  Minkowski's metric has  nothing
to  do  with the velocity of the particle. However, since $v<<1$, we
can use Newtonian coordinates  (with $x^4 =t$, $t$ being the
Newtonian time),   so that  the spatial components of the geodesic
equations become \be
 \frac{d^2
x^{i}}{dt^2} =
 -\Gamma^{i}_{\mu\nu} \frac{d x^{\mu}}{dt}
 \frac{dx^{\nu}}{d t} -2\Gamma^{i}_{\mu 4} \frac{d x^{\mu}}{d t}=
 -\Gamma_{44}^{i} = -\frac{1}{2}\delta h_{44,i}, \;\;\; i,j =1..3 \label{eq:geodesic}
 \ee
At this point  a scalar field $\varphi$  may  be  defined   such
that
 \be
 \frac{d^2 x^{i}}{dt^2} =
   -\frac{\partial\varphi}{\partial x^{i}}
 \label{eq:newton}
\ee Comparing \rf{geodesic} and \rf{newton}
 we obtain an equation to determine the field $\varphi$:
 \be
\frac{\partial\varphi}{\partial x^{i}} =
-\frac{1}{2}\delta_{ij}\frac{\partial\delta h_{44}}{\partial x_j}
\label{eq:smallphi}\ee
As the particle continues its fall, the gravitational pull
continuously builds up by small increments of the metric  as  in
 $$ g_{\mu\nu} \approx \eta_{\mu\nu} +\delta h_{\mu\nu}+ (\delta
 h_{\mu\nu})^2 +\cdots$$
Actually, there is no way to stop this process  without applying an
external force. Thus,  equation \rf{smallphi} can be integrated
along the  geodesic, where  $\delta h_{\mu\nu}$ continuously
increase up to a finite value $h_{\mu\nu} $, leading to the scalar
gravitational  potential (often referred to as the nearly Newtonian
potential,  not to be confused  with the  post  Newtonian
approximations):
 \be
 \varphi= -\frac{1}{2}\int_{0}^{h_{44}} d(\delta h_{44})=
-\frac{1}{2}(1+g_{44}) \label{eq:NND} \ee Notice  that  except at
the  beginning of the free fall a  weak gravitational  field  was
not  imposed.  Here $g_{44}$ is obtained from an exact solution of
Einstein's equations,  so that  at the end  all   metric
  components
$g_{\mu\nu}$  contribute  to \rf{NND}.
 In the following we  exemplify this application of  \rf{NND} to the motion of stars in galaxies and clusters.

\%subsubsection{Rotation  Velocity  curves   in Galaxies}

The gravitational  field of  a   simple   disk galaxy   model  can
be  obtained  from the cylindrically symmetric   Weyl  metric,
expressed in cylindrical coordinates  $(r,z,\theta,t)$ as
\cite{Weyl}:
\begin{equation} dS^{2}
=e^{2(\lambda-\sigma)}dr^{2} + r^{2}e^{-2\sigma}d\varphi^{2}
+e^{2(\lambda-\sigma)}dz^{2} - e^{2\sigma}dt^{2} \label{eq:weyl}
\end{equation}
 where $\lambda =\lambda{(r,z)}$ and $\sigma =\sigma{(r,z)}$.
 As   shown in  \cite{Rosen,Zipoy},  the Weyl cylindrically
symmetric solution  is diffeomorphic to the Schwarzschild solution.
Replacing    the Weyl   metric  with  these conditions in   \rf{BE1}
and  \rf{cons} we  obtain
\begin{eqnarray}
 && - \lambda_{,r} + r\sigma_{,r}^{2} -r
\sigma_{,z}^{2} =0
\label{eq:first}\\
 &&-\sigma_{,r}-r\sigma_{,rr}- r\sigma_{,zz}=0 \label{eq:second}\\
&& \lambda_{,rr} +\lambda_{,zz} +\sigma_{,r}^2 + \sigma_{,z}^2
=0\label{eq:third}\\ && 2r\sigma_{,r}\sigma_{,z}
=\lambda_{,z}\label{eq:fourth}
\end{eqnarray}
The   cylinder  solution
is  diffeomorphic  to  a Schwarzschild's  solution\footnote{ This  is  a fine example of the
equivalence problem in general relativity: How do we know that two
solutions of Einstein's equations, written in different coordinates,
describe the same gravitational field? The  answer is given
by the  application of Cartan's equivalence problem   to general
relativity. It shows that the Riemann tensors and  their covariant
derivatives up to the seventh order   must be equal \cite{MacCallum}.}.

  In the following  we  apply    this  solution   to  find  a  the geodesic  motion  of   slowly   free  falling  star under the gravitational  field    in  the galactic plane. Since the  slow motion geodesic  equation
\rf{newton}  is not  invariant under diffeomorphisms,    we  may
consider three separate stages  separately, in accordance with the
symmetry of the gravitational field which is effective at the
current position of the  star:

\emph{(A) {\textbf{The star is far away from the galaxy}}}:\\
In this case,    the gravitational field  of the galaxy
 is weak, like that of a point source. Therefore, the predominant
gravitational field is given by the  exterior Schwarzschild
solution, as seen from a large distance  from the galaxy  nucleus.
In spherical coordinates, we   can  write  the  metric component
$g_{44} =-(1 -2M/r)$, and \rf{NND} gives
$$\varphi\rfloor_{Newton} =-M/r, \;\;\; r>>r_0$$
which is   equivalent to the Newtonian gravitational potential
produced by a distant mass $M$,   determined by the  Newtonian (weak
field) limit.  In such situation \rf{NND} agrees with all estimates
resulting from the Newtonian gravitational theory for a spherically
symmetric dark matter halo, producing the same rotation velocity
curves.\\
\emph{(B) {\textbf{The star is close to the galaxy's disk}}}:\\
 When the star is close to the galaxy, in the disk plane,
the predominant dark matter gravitational field can be   simulated
by
 the vacuum cylindrically symmetric Weyl   metric,
satisfying the vacuum  Einstein's  equations \rf{first} to
\rf{fourth}. However, to  configure  a disk galaxy    we need to
apply  the  condition  that Weyl's  cylinder   has a thickness  that
is  much  smaller than  its  radius: $h<<r_0$.  With this  condition
we can  no longer  apply   the diffeomorphism  invariance of general
relativity,  so that   the  solution must be  written in
cylindrical  coordinates.

The  disk-symmetry  condition  can be written  as  $ z\in [-h_0/2,
h_0/2],\;\;\; \mbox{for} \;\;\; r\in [0,r_0],\;\;\; h_0<<r_0 $, so
that    the functions $\sigma(r,z)$ and $\lambda(r,z)$ may be
expanded around $z=0$ as
  \begin{eqnarray*}
&&\phantom{x}\hspace{-5mm}\sigma(r,z)
=\sigma(r,0)+ z a(r) + \cdots \\
 &&\phantom{x}\hspace{-5mm}\lambda(r,z)
=\lambda(r,0) + z b(r) + \cdots
 \end{eqnarray*}
 where we have denoted
$ a(r)= \left.   \frac{\partial \sigma(r,z)}{\partial
z}\right\rfloor_{z=0} \;\; \mbox{and}\;\;\,b(r)=\left.
\frac{\partial \lambda(r,z)}{\partial z}\right\rfloor_{z=0} $.
Neglecting the  higher order terms, it follows that equations
\rf{second} and \rf{fourth} become a simple system of equations on
$\sigma$, with general solution $\sigma(r,z)=\frac{K}{2}\ln r + c_2
(z)$ where $c_2(z)$ is an r-integration constant and where we have
denoted $ K ={b(r)}/{a(r)} $.
 Derivation of $\sigma$
 with respect to $z$ gives $c_2(z) =a(r)z + c_0$, but since
  $c_2$
 does not depend on $r$, it follows that
  $a(r)$ must be a constant $a_0$. By similar arguments we find that
  $b (r) =b_0 =$constant, so that $K = K_0= b_0/a_0$ is also a constant.
Replacing these results in \rf{first} and \rf{third}, we again
obtain another simple solvable system of equations in $\lambda$, so
that the solution of the vacuum Einstein's equations for the Weyl
disk is
  \begin{eqnarray}
&&\sigma(r,z)=\frac{K_0}{2}\ln r + a_0 z + c_0 \label{eq:sigma}\\
 &&\lambda(r,z) =
\frac{K_0^2 }{2}\ln r - a_0 \frac{r^{2}}{2}+b_0 z + d_0
\label{eq:lambda}
 \end{eqnarray}
 where $c_0$ and $d_0$ are again  integration constants.
 From \rf{sigma} we
obtain $g_{44}= -e^{2\sigma}=-e^{2\frac{K_0}{2}\ln r} e^{2a_0 z}
e^{2c_0}$. Therefore,  for a star near the galaxy
 in the galaxy plane $z=0$, in the region between
the nucleus  radius $r_c$ and the disk radius $r_0$    \rf{NND}   is
 \be
 \varphi\rfloor_{disk}= -\frac{1}{2}(1+ g_{44})\rfloor_{z=0} = -\frac{1}{2}(1
-e^{2c_0} r^{K_0} ), \;\;\; r_{c}<r<r_{0}
 \label{eq:NNDdisk}
 \ee
Here  we cannot make use of the Newtonian limit to determine the
integration constant $e^{2c_0}$ in \rf{NNDdisk}, because we do not
have the same symmetry and the same boundary conditions appropriate
for the Newtonian gravitational field. Instead,  we  may   compare
\rf{NNDdisk}with the local Newtonian potential produced by a   disk
of   visible mass M, suggesting that the above integration constant
can be written as proportional to the to the total baryonic mass
$M_b$ of the galaxy. Thus, in units G=c=1 we set $
e^{2c_0}K_0/2=\beta_0 M_b$, where $\beta_0$ is a mass scale factor,
proportional to the estimated visible mass of the galaxy
\cite{Salucci}.

The rotation velocity of a test particle under \rf{NND} is
$v=\omega_0 r$, $\omega_0=$ constant,
 obtained by comparing the radial force
$\frac{v^2}{r}\hat{r}$, with $ -\frac{\partial \varphi}{\partial
r}\hat{r} $, so that the velocity is given by $v =
\sqrt{|r\frac{\partial\varphi}{\partial r}|}$. In the particular
case of the disk, using \rf{NNDdisk} we obtain
 \be
v(r)=\sqrt{|\beta_0 M_b r^{K_0}|}, \;\;\; r_c<r<r_0
\label{eq:velocity}
 \ee
Here   $K_0$  represents  the ratio between  the coefficients  $a(r)$ and  $b(r)$ of the
expansion  of  the  metric functions  $\sigma$ and  $\lambda$
respectively.  In the  considered  linear  expansion in \rf{sigma}
and  \rf{lambda},   $K_0$  can be   adjusted  experimentally. In
practice it can be     scaled by  a constant,  so as  to belong to
interval between $0$  and  $1$,  corresponding to the  galaxy
nucleus radius  where the  spherical symmetry applies,   and   to
the  distance where the disk symmetry applies.  The extreme  values
in that  interval  are   excluded:  The value $K_0= 0$ is   excluded
because for that value the potential \rf{NNDdisk} does not exert any
force. The value $K_0=1$ is also excluded because it gives velocity
proportional to $\sqrt{r}$ which is not experimentally verified.

Using the    minimum  square   root curve  fitting, figures 2, 3 and
4 show the velocities near the disk, calculated with \rf{velocity}
for some known galaxies (red dots) for $r$ larger than the estimated
core radius where the disk symmetry applies. The values of the two
free parameters $K_0$ and $\beta_0$ in each case were determined by
the minimum squares numerical iteration method, extracted from known
experimental data \cite{Sanders}. For comparison purposes, we
included  the black error bars showing  the measured velocities and
the green triangles corresponding  to  the Newtonian  predictions.
  \begin{figure}[!h]
 \begin{center}
\includegraphics[width=6cm]{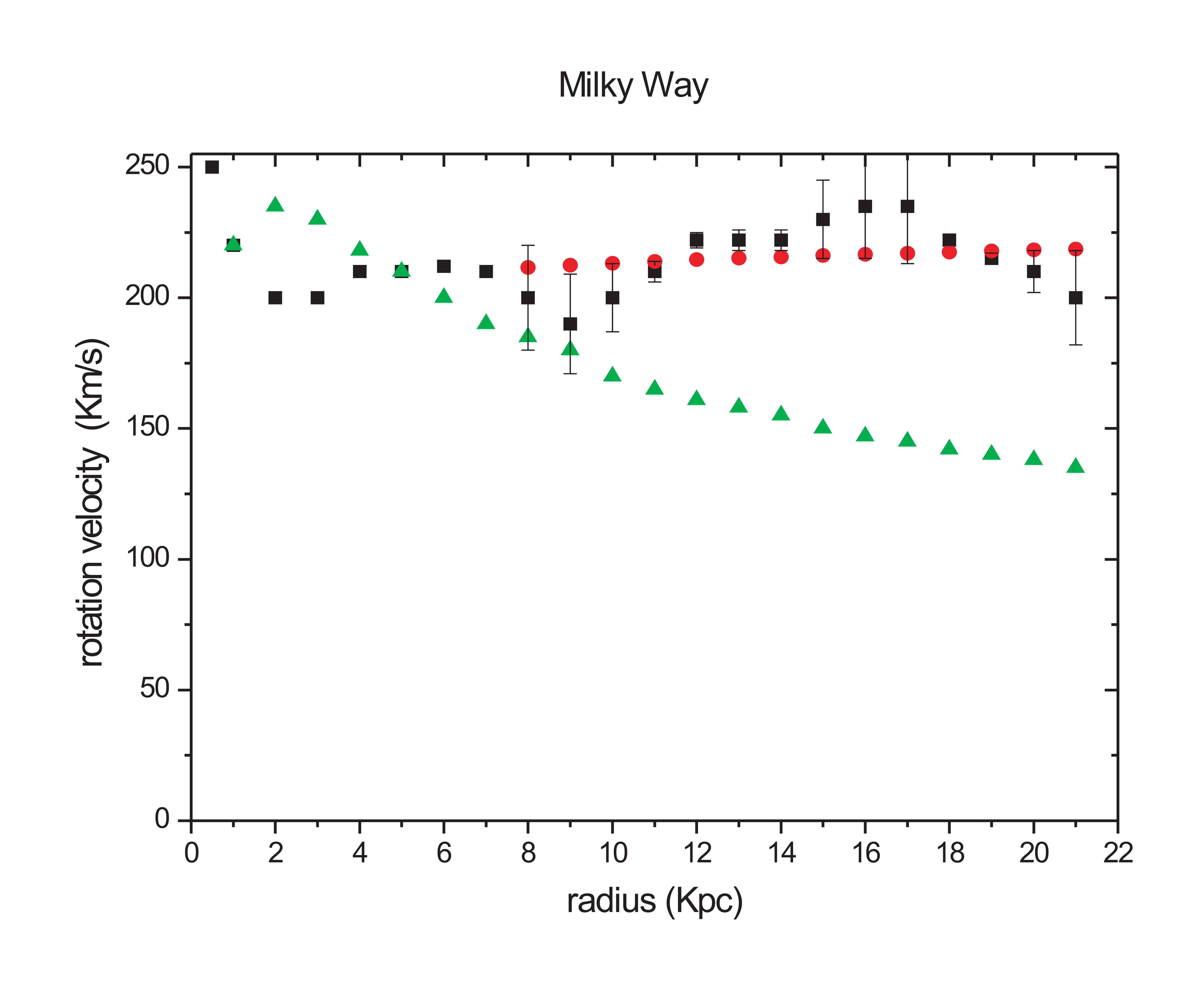} \caption{Via Lactea
with $M_b = 1.99\times 10^{41}$Kg, $R=8,5$Kpc, $K_0=0.0682$ and
$\beta_0=0.0901 $. } \end{center}
\end{figure}
 \begin{figure}[!h]
\begin{center} \includegraphics[width=6cm]{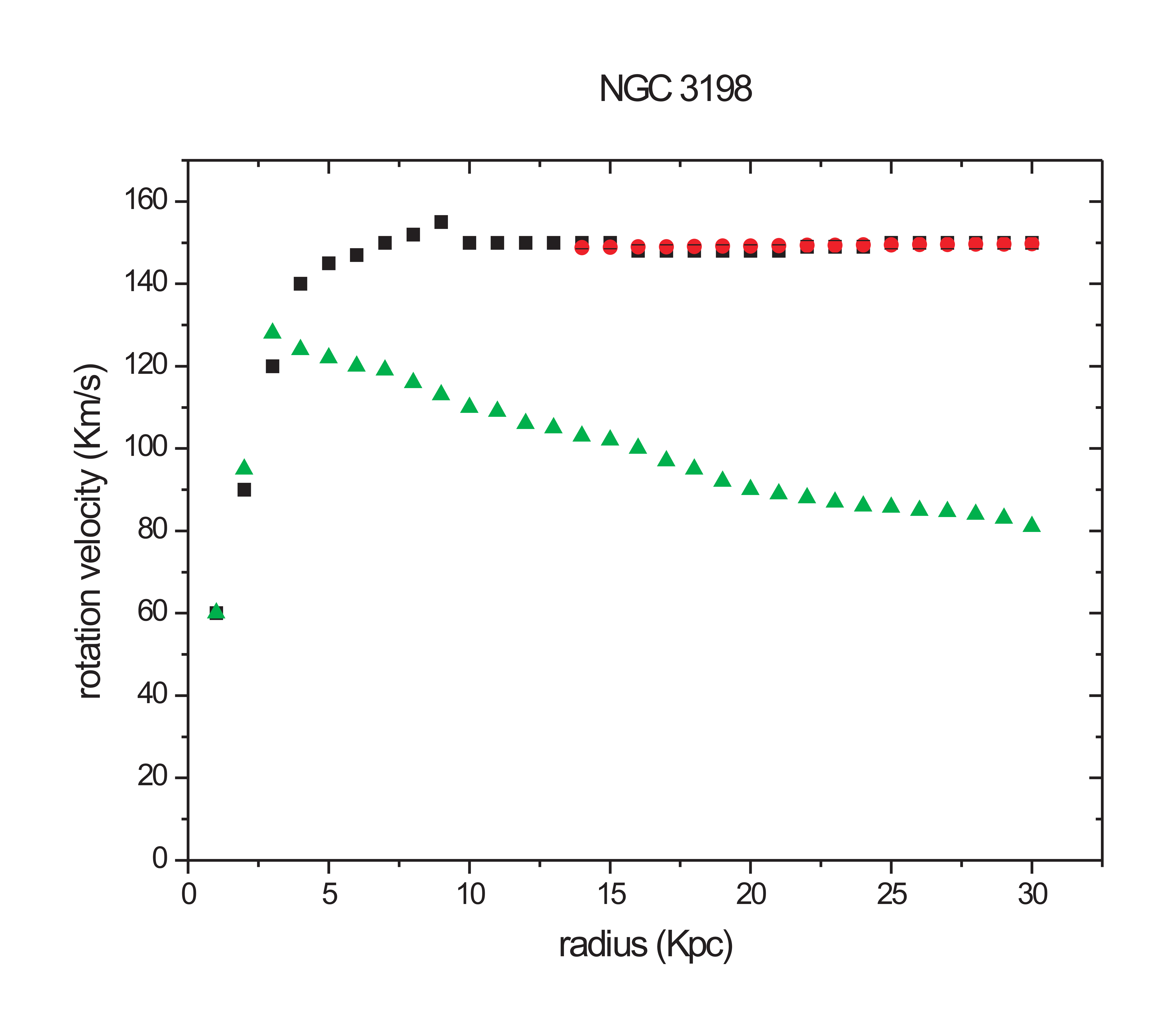}
\caption{NGC3198, with $M_b = 1.19\times 10^{40}$Kg, $R=16,6$Kpc,
$K_0=0.0162$ and $\beta_0= 0.8224$ } \end{center}
 \end{figure}
\begin{figure}[!h] \begin{center}
\includegraphics[width=6cm]{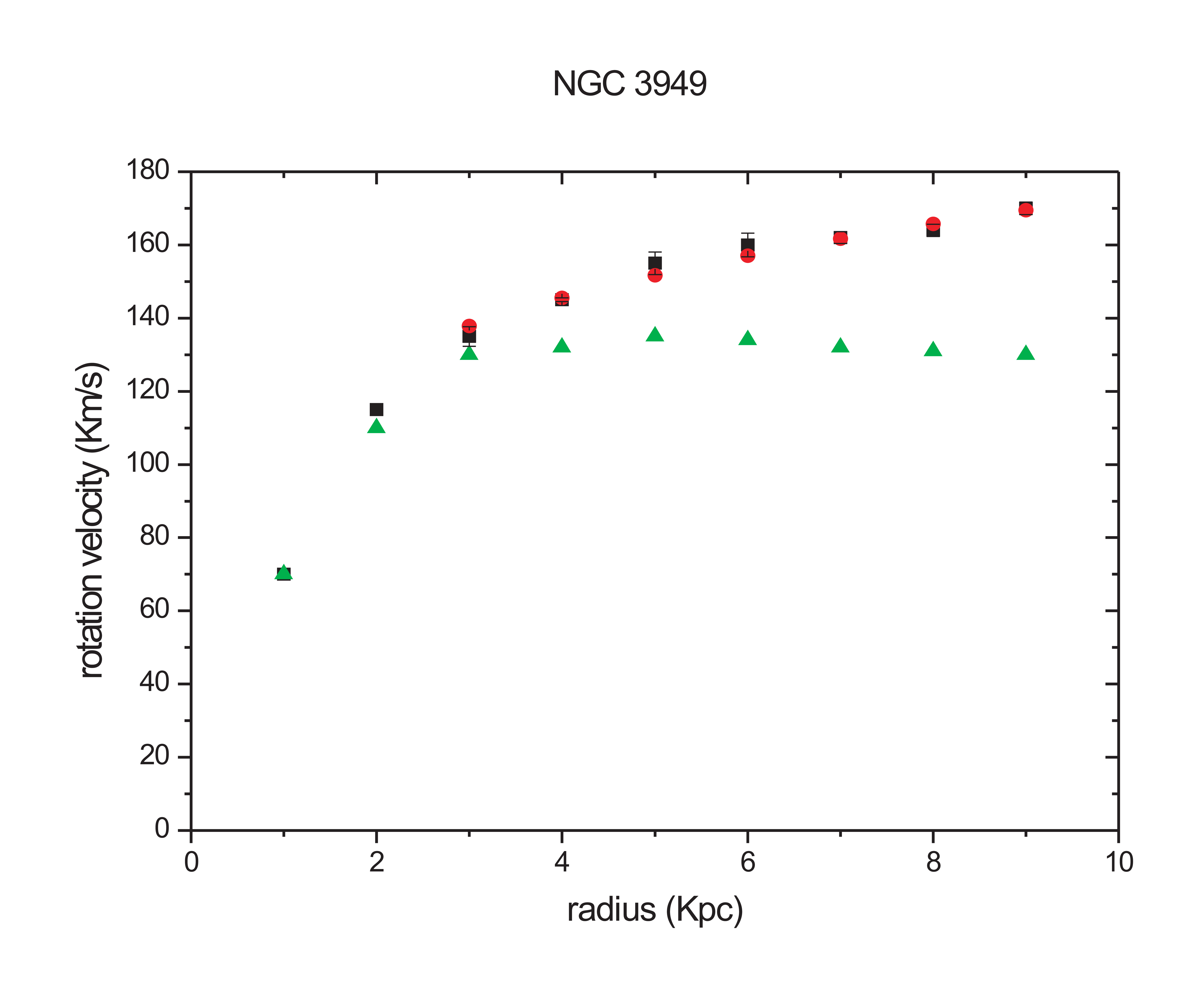} \caption{NGC3949, with $M_b
= 4.97\times 10^{39}$Kg, $R=6$Kpc, $K_0= 0.3766$ and
$\beta_0=1.1657$}
\end{center}
 \end{figure}


\emph{(C) {\textbf{The star is near the galaxy's  nucleus}}}:\\
When the star reach the nucleus of the galaxy, the observed velocity
is still small,  but the gravitational  field  is  likely   to be
strong. A  simple  configuration  of the  gravitational field
 acting on the star  can be  obtained  if  we neglect
  the rotation of the  galaxy's  nucleus, so that we  obtain a
 spherically symmetric gravitational  field,  described by the
 by the Schwarzschild solution of  \rf{EEvac} with
  $g_{44} =-(1 -2M/r)$. Then we obtain  from \rf{NND}
\be
 \phi\rfloor_{Nucleus} =-M/r, \;\;\; r \le r_c    \label{eq:core}
\ee
 and  the star   rotation  velocity  is   given by
  $v(r) = \sqrt{{M}/{r}}$, were now $M$ is
 associated with the   Schwarzschild mass of the Nucleus.
 Thus, the velocity   looks exactly like  the one  in  the Newtonian
 theory,  with the exception that the gravitational field is not
 necessarily  weak. This  is  a  consequence of the assumed
  metric  symmetry  at  the  galaxy's  nucleus.


Two    recent observations   of  colliding   cluster dark matter
halos have provided  new insights on the dark matter  gravity issue:
One of them is  the  bullet cluster 1E0657-558,  showing    the
motion of a sonic boom   effect  in  the intercluster   plasma  with
velocity $\sim 4700$km/s, visible  through x-ray astronomy
\cite{Clowe}. Under the   assumption that the gravitational field of
the  two  clusters  is  Newtonian and the existence of gravitational
field of dark matter halos in each  cluster, the (linear)
superposition of the two  dark matter gravitational fields produce a
center of mass of the system which coincides with the observed
position of the plasma bullet.  This  has been claimed to be  an
observational evidence for the existence of dark matter.

 On the other hand,  using \rf{NND} instead of the Newtonian gravity,
 and  admitting that the   gravitational field of  each  cluster  is
   spherically  symmetric,  we  have
the gravitational  field  of the spherically symmetric dark matter
halos would be  given by two separate Schwarzschild solutions. Then
 the  action of each of these  gravitational field on the plasma
 can be taken  as in  \rf{core}, which resembles
  the Newtonian potential,  producing the same center of mass as described  in
Newtonian theory. However,  such   superposition of two
Schwarzschild's solution  is  only a  crude  approximation  to
compare  with  the dark matter  halos. To be more precise, either we
consider  a two body   problem  solution  of  \rf{EEvac},  or  else
we  consider  the  cluster  collision as   a  Nash's perturbation
process. In the latter  case  we  may start  with an  embedded
spherically symmetric  cluster   which is  perturbed  by  the
extrinsic curvature generated   by the  second  cluster  according
to  \rf{York}.  This  is   easier  said  than   done,   because the
differentiable embedding of  a  Schwarzschild  solution requires six
dimensions.  Nonetheless,  in principle this  solution   can be
calculated and tested.

The  second  observation  is   that of the Abel 520  cluster
MS0451+02,   showing  again two  colliding    "dark matter  halos",
where at least one of them  do  not seem  to be  anchored to a
baryonic  structure.  The  most  immediate  explanation is that this
"pure  dark matter halos"  may be an  evidence  of  a  non-linear
effect  of the  dark matter gravitational  field,  which  does not
agree with the  Newtonian  gravitational  field  assumption. So,  in
a  sense  this  observation backs  up the  hypothesis  of   a
non-linear  alternative  gravitational  theory  at the
galaxy/cluster  scale  of  observations.

In particular  it is possible  to  explain  the existence of  a
gravitational   effect  which is  not  anchored  to a baryonic
source  as  a  solution of  \rf{EEvac},  as    a  for    example a
vacuum Schwarzschild solution or  a  Weyl disk  solution, which may
act  as  a perturbation  to  another   cluster,  again  applying
\rf{York} to  find  the final gravitational field of the system. The
 result  may be  compared    with the motion of the
x-ray observed plasma structure. However,  we are still pending on
further details on the Abel 520 collision.

\section{Summary}

Gravitational  perturbation theory   has become an   essential tool  for  the  explanation of  the   formation of  large structures   in the  universe.
Here  we have   presented  an application of Nash's  theorem  on perturbations of  geometries  to  the  formation  of space-time structures  and  to explain the
local effects usually  attributed to  dark matter.
In a   brief  justification  we  have argued that  the  theorem  improves  Riemann's  geometry   in the sense  that   it   replaces the   somewhat
absolute notion of  Riemann curvature,  by a  relative notion of  curvature  with  respect  to  the Riemann tensor of the bulk
 defined  by  the Einstein-Hilbert principle.

The  relevant  detail in  Nash's theorem,  is that it provides   a  mathematically sound and coordinate gauge free way to
construct any Riemannian  geometry, and  in particular any
space-time  structures, by a  continuous sequence of infinitesimal
perturbations along the  extra dimensions of the  bulk space,
generated by the  extrinsic  curvature.

The four-dimensionality of  space-time  is regarded here as an
experimental fact, related to the  symmetry properties of Maxwell's
equations  or, more  generally of the   gauge theories of the
standard  model.  Thus,  any matter that interacts  with  gauge
fields,  including  the   observers, must remain  confined   to
four-dimensions. However,   in accordance with  Nash's   theorem,
gravity as  represented  by a  metric cannot be  confined because it is
 perturbable   along the extra dimensions.

In a  first cosmological application of   that theorem  we  have  compared  our results with the present  observational  data. Starting  with the FLRW
cosmological model  embedded in the  five-dimensional
 deSitter bulk, and  applying  equations  \rf{BE1} and \rf{BE2}.
 We  have  found that Friedman's equation is necessarily  modified   by the presence of the extrinsic curvature. We have shown  that  this  modification is
consistent   with the observed  acceleration  of the universe, including  with the observed   power  spectrum of the CMBR.

In principle  the   local   effects  of  dark matter   gravitation
should have the same  explanation,  although they  have  different   observational   aspects.  The simplest   case is that of the   rotation  curves  in galaxies  and  galaxy clusters,  which originated the   dark matter issue.  In this  case,   there is  no  evidence  that the gravitational perturbation process   is   still active, except perhaps  in  young galaxies.  Therefore,  we  have   considered  the case  where  Nash's perturbations  vanish,   obtaining  the vacuum   Einstein's  equations,  applied  to   the  metric  with an  specific   symmetry.
Instead  of the Newtonian  potential  we  have applied  the  geodesic  equations  for  slow  motion.  Using  the  Weyl metric  to  simulate  a disk galaxy, we  have    compared  the result  with   some  known  cases.

 More recently  the observations  of  merging clusters  have provided  additional information on the dark matter issue.  The slow  motion of the  plasma  substructures    can  be    handled  by the same  equations  but    it  appears to us  that the correct   formulation of the problem  should be  made  with
Nash's  perturbative  analysis.

\end{document}